\documentstyle[art12,bezier]{article}

\begin{document}

\renewcommand{\thefootnote}{\fnsymbol{footnote}}

\thispagestyle{empty}

\hfill \parbox{45mm}{
{MPI-PhT/95-6} \par
January 1995}

\vspace*{15mm}

\begin{center}
{\LARGE General estimate for the graviton lifetime.}

\vspace{22mm}

{\large Giovanni Modanese}%
\footnote{A. Von Humboldt Fellow.}

\medskip

{\em Max-Planck-Institut f\"ur Physik \par
Werner-Heisenberg-Institut \par
F\"ohringer Ring 6, D 80805 M\"unchen (Germany)}

\bigskip \bigskip

\end{center}

\vspace*{10mm}

\begin{abstract}
By means of general kinematical arguments, the lifetime $\tau$ of a
graviton of energy $E$ for decay into gravitons is found to have the
form $\tau^{-1} = \frac{1}{EG} \sum_{j=1,2,...} c_j (\Lambda G)^j$.
Some recent, preliminary
results of non perturbative simplicial quantum gravity are then
employed to estimate the effective values of $G$ and $\Lambda G$.
It turns out that a short lifetime of the graviton cannot be
excluded.

\bigskip \bigskip

\end{abstract}

The recent advances of non perturbative simplicial quantum gravity
\cite{h1,h2,mod} have made possible a computation
of the scale dependence of the Newton constant $G$ and
especially of the adimensional product $\Lambda G$.
In this paper we consider a process, the decay of the graviton, whose
probability admits a very simple expression in terms of $G$ and $\Lambda G$,
and we give a general estimate for the lifetime.

Let us begin with some kinematical considerations.
The decay of a massive particle into $n$ product particles
is described by the well known formula for the lifetime $\tau$ in the
rest system of the particle, namely
\begin{equation}
  \tau^{-1} = \frac{1}{2M} \int \prod_{i=1...n}
  \frac{d^3 p_i}{(2\pi)^3 2E_i} \, \delta(M-\sum_i E_i)
  \, \delta^3(\sum_i {\bf p}_i) \, |T|^2 ,
\end{equation}
where $T$ is the quantum amplitude for the process and $M$ is the mass
of the decaying particle. In a reference system in which the particle
moves with some velocity $v$, its lifetime $\tau$ appears to be larger
by the factor $\gamma=(1-\beta^2)^{-1/2}$ ($\beta=v/c$; unless
otherwise specified, we put $c=\hbar=1$).

If the decaying particle is massless, there exists no rest frame for it,
but a very simple formula for the lifetime can be obtained anyway.
Let us suppose that in some reference system the four-momentum of the
particle is
\begin{equation}
  p = (E, \, E, \, 0, \, 0),
\label{dk}
\end{equation}
that is, the particle moves along $x$, from the left to the right,
with energy $E$. Consider a Lorentz boost along the $x$ axis,
namely of the form
\begin{equation}
  L(\beta) = \left(
  \begin{array}{cccc}
  \gamma & - \beta \gamma & 0 & \ 0 \\
  - \beta \gamma & \gamma & 0 & \ 0 \\
  0 & 0 & 1 & \ 0 \\
  0 & 0 & 0 & \ 1
  \end{array}
  \right) .
\label{ke}
\end{equation}

For any vector $V$, we have in the boosted system $V'=L(\beta)V$,
that is
\begin{eqnarray}
  & {V_0}' = \gamma (V_0 - \beta V_x) ; \nonumber \\
  & {V_x}' = \gamma (V_x - \beta V_0) ; \nonumber \\
  & {V_y}' = V_y; \qquad {V_z}' = V_z.
\label{sn}
\end{eqnarray}
At the time $t=0$, the origins of the two systems coincide. Suppose
now that the massless particle is produced at $t=0$ with the
four-momentum $p$ above (eq.\ (\ref{dk})) and its decay is observed
in the unprimed reference system at a time $t=\tau$, that is, at a
coordinate $x=\tau$. Using (\ref{sn}) to transform $E$ and $\tau$
in the primed reference system one sees immediately that
\begin{equation}
  \frac{E'}{\tau'} = \frac{E}{\tau} ,
\label{ks}
\end{equation}
that is, the lifetime of a massless particle in any reference system is
proportional to its energy:
\begin{equation}
  \tau = \xi E .
\label{sw}
\end{equation}
The constant $\xi$ depends on the dynamics and has dimension
$[length]^2$. The simple formula (\ref{sw}) is due entirely to
Lorentz invariance.

Another consequence of Lorentz invariance is the following. If we make
a boost in the positive $x$ direction with $\beta$ close to $1$, say
$\beta=1-\varepsilon$, the energy $E$ of a massless particle with
four-momentum (\ref{dk}) reduces to
$E'\simeq \sqrt{\varepsilon/2} \, E$. (If the
boost is in the direction opposite to the motion of the particle, we
have instead $E' \simeq E/\sqrt{\varepsilon/2}$.) It follows that a
massless particle can obviously decay only in massless particles;
otherwise, in the reference systems in which the energy of the
initial massless
particle is smaller than the total mass of the products, the process
would be impossible.

Moreover, if the momentum of the decaying particle points in a certain
direction, like in (\ref{dk}), the momenta of all product particles
must be parallel to the same direction (Fig.\ 1). For, if the momenta
of the products would have some components (with vanishing sum) in an
orthogonal direction, then performing a boost along $x$ such that the
initial energy becomes $E'=\sqrt{\varepsilon/2} E$, the orthogonal
components would be not affected, thus spoiling energy conservation
(because $E'= \sqrt{{p_x'}^2+{p_y'}^2+{p_z'}^2}$). Since all
momenta are parallel, it also follows that if one particle with some
helicity $h$ decays in $n$ particles with helicities $\pm h$, in
order to conserve the angular momentum $n$ must be odd.

Although the considerations above are quite elementary, they are not
to be found in the literature, probably because this decay process
is not relevant for photons or massless gauge bosons. Now we can
understand why. We note first that in both cases perturbation theory
gives an adequate treatment. So, the photon does not decay
into three photons through the ``fermion square loop'' which gives
a small photon/photon scattering amplitude in QED, because it is
straightforward to check that for the configuration of momenta and
helicities described above the transition amplitude vanishes. This
conclusion can be generalized to the $n$-photons amplitude [details
will appear elsewhere]. Also, if one considers the decay amplitude
of a massless Yang-Mills gauge boson into three bosons through the
quartic coupling, one easily finds one more time that the amplitude
vanishes for the described configuration of helicities and momenta.

A physically interesting case in which a massless particle is allowed
in principle to decay into other massless particles is that of the graviton.
It is known that the non linearity of Einstein equations causes
in gravitational waves an interaction between different frequency
modes, a phenomenon known in the theory of non-linear differential
equations as the ``energy cascade''. But it is not clear yet
how much relevant this phenomenon can be (see \cite{efr} and references).

In our opinion, it is more appropriate to investigate
it quantum mechanically, that is, as a decay of the graviton
into three or more gravitons. Let us apply eq.\ (\ref{sw}) to this case.
We stay at a very general level, and assume only that the two
dimensionful constants $G$ (Newton constant) and $\Lambda$
(cosmological constant) enter in the dynamics of the process. From
dimensional arguments it is easily found that the reciprocal of the
lifetime is given by
\begin{equation}
  \tau^{-1} = \frac{1}{EG} \sum_{j=1,2,...} c_j (\Lambda G)^j ,
\label{za}
\end{equation}
where $c_j$ denotes an unknown adimensional coefficient.

We have excluded the values $j \leq 0$, because the decay probability
is proportional to $\hbar^j$, as can be checked restoring
the constants $\hbar$ and $c$. Of course, the dimensional estimate does
not allow us to pick up the relevant $j$. This requires a knowledge of
the dynamics of quantum gravity which is still far from being obtained.

If we substitute in (\ref{za}) the observed values $G \sim 10^{-66} \ cm^2$,
$\Lambda G < 10^{-120}$, and remind that in natural units
$1 \ eV \sim 5 \cdot 10^4 \ cm^{-1}$, we find that in any case
the lifetime is very long
\footnote{The situation is different if we are concerned with early
cosmology. In the inflationary model, for instance, the parameter
$G \Lambda$ can be at some stage of the order $10^{-12}$.}.
But we would like, in the following of the paper, to expose a further
point of considerable interest: while the effective value of $\Lambda G$
observed at astronomical distances is certainly very small, in the
quantum theory this quantity depends on the length scale and can turn
out to be much larger, possibly leading to an observable lifetime.

As a model for the quantum theory, we consider the euclidean functional
integral approach of Hamber and Williams \cite{h1}. This technique
has been developed for almost 10 years now, and it is under many
regards pretty successful. It is based on the numerical
non perturbative evaluation of the discretized partition function
\begin{equation}
  Z = \int_{\rm Geometries} d[g] \, e^{-S[g]}
\label{xq}
\end{equation}
where the geometries are described by Regge simplicial manifolds.
The action in (\ref{xq}) has the form
\begin{equation}
  S = \int d^4x \, \sqrt{g} \left( \lambda - kR +
  \frac{1}{4} a R_{\mu \nu \rho \sigma} R^{\mu \nu \rho \sigma}
  \right) .
\label{op}
\end{equation}

The constants $k$, $\lambda$ are clearly related, as ``bare'' quantities,
to $G$ and $\Lambda$: $k$ corresponds to $1/8\pi G$ and $\lambda$ to
$\Lambda/8\pi G$. It is important, however, to keep distinct the physical
values $G$ and $\Lambda$ from $k$ and $\lambda$. The latter
are entered as parameters at the
beginning, and then a second order transition point for the
statistical system described by $Z$ is found by Montecarlo simulation.
Actually, there is a line of transition, for one can also vary the
adimensional parameter $a$, which does not have ``macroscopic'' counterpart.
On this line in the parameter space the theory admits a continuum limit.
Unlike in perturbation theory, where a flat background is
introduced by hand, the flat space appears here dynamically:
the average value of the curvature vanishes on the transition line,
separating a ``smooth phase'', with small negative curvature, from a
``rough'', unphysical phase with large positive curvature.
In this way the effective, long scale cosmological constant
\begin{equation}
  \Lambda \sim \frac{\left< \int \sqrt{g} R \right>}
  {\left< \int \sqrt{g} \right>}
\end{equation}
vanishes in the continuum quantum theory, confirming a recent
analytical computation of Greensite (\cite{gre}; compare also the
praevious works of Coleman \cite{col}, Taylor and Veneziano \cite{ven}
and the review of Weinberg \cite{wei}).

Using the preliminary non perturbative results of \cite{h2},
we now make these considerations more quantitative. In the
action (\ref{op}) the parameter $\lambda$ has the task of fixing the
length scale, so that every dimensional quantity is proportional
to some power of $\lambda$. In the following, we denote with a tilde
quantities which are pure numbers. For instance, the mean length
$l_0=\sqrt{\left< l^2 \right>}$ of the edges of the dynamical lattice
is given by $l_0=\tilde{l}_0 \lambda^{-1/4}$.
With $a=0$, it is found $\tilde{l}_0 \simeq 2.36$.

The length scale $\lambda$, however, is not arbitrary like a cutoff in usual
lattice field theory. Since the scale dependence of $G$ is expected
to be very weak at the continuum limit, $\lambda$ can be fixed by the
condition that the parameter $(8 \pi k_{critical})^{-1}$ coincides with
the physical Newton constant $G$. The latter is extracted from the
potential energy $U$ of two sources of mass $M$, namely
\begin{equation}
  G = \frac{Ur}{M^2} = \frac{\tilde{U}_N \tilde{l}_0 N}{\tilde{M}^2}
  \lambda^{-1/2} .
\end{equation}
In this formula, $N$ is the number of lattice spacings considered
\cite{h2,mod}. On the Regge lattice the two sources can be placed
at a mutual distance which is a multiple of the lattice
spacing; if $U_N$ is the corresponding potential, the quantity $N U_N$
should be constant, but for small values of $N$ it can be sensitive to
lattice artifacts and for large values of $N$ to volume effects.
So $N$ must usually be chosen in an appropriate range.

It is very encouraging that already in the preliminary simulation the
values of $(8 \pi k_{critical})^{-1}$ and $G$ are comparable. Both turn out
to be, in $\lambda$ units, of the order of unity; for instance, it is
found that
\begin{equation}
  k_{critical} \simeq 0.06 \, \lambda^{1/2} .
\label{vf}
\end{equation}
Eq.\ (\ref{vf}) expresses the fact that the natural length scale of the
lattice, $\lambda^{-1/4}$, is effectively of the order of Planck length, as
expected on physical grounds.

On the other hand, the effective cosmological constant has a strong
dependence on the energy scale $\mu$. Very close to the phase transition
the adimensional quantity $\Lambda G$ should behave like
\begin{equation}
  (\Lambda G)(\mu) \sim \tilde{k}_c^{-2} (l_0 \mu)^\gamma ,
\label{hg}
\end{equation}
where the critical exponent $\gamma$ is estimated to be $\gamma \simeq 1.54$.
Thus, returning to (\ref{za}), we must substitute (\ref{hg}) in the
terms of the series. The energy scale $\mu$ for the decay process cannot be
exactly precised at this stage
\footnote{It is connected to the spatial size of the wave function
of the graviton, and/or to the features of the measuring apparatus.
$\mu$ is not related to the energy $E$ of the graviton, because
if the constant $\xi$ in (\ref{sw}) would depend on $E$, say $\xi=\xi(E)$,
then it would become $\xi(E')$ in a boosted system; but $\xi(E')=\xi(E)$,
due to (\ref{ks}).}
, but it is roughly of the order of the ``laboratory'' scale
$\mu^{-1} \sim 1 \ cm$.

It should be recalled at this point that the
measurement which leads to the macroscopic limit $\Lambda G < 10^{-120}$
corresponds in practice to the astronomical observation that the radius
of curvature of the known universe is larger than $\sim 10^{30} \ cm$. For
this measurement, the scale $\mu$ to be inserted in (\ref{hg}) is then much
smaller than the laboratory scale.

Since $l_0 \sim L_{Planck}$, the product $(l_0 \mu)$ at laboratory energies
is of the order of $L_{Planck}/cm=\sqrt{G}/cm$ and we have from (\ref{za})
and (\ref{hg})
\begin{equation}
  \tau^{-1} \simeq \frac{1}{EG} \sum_{j=1,2,...} c_j
  (\sqrt{G}/cm )^{j \gamma} .
\end{equation}

The dominant term is that with $j=1$, which gives
$(\tau/E) \simeq (c_1)^{-1} (G/cm^2)^{0.23} \ cm^2$. This corresponds to a
quite short lifetime for the graviton. Of course, it is still
just a rough estimate, due to the uncertainty on $\mu$ and $\gamma$
(we may expect that in order to get consistently
$\Lambda G < 10^{-120}$ for $\mu^{-1} \sim 10^{30} \ cm $,
the exponent $\gamma$ should result to be eventually closer to 2).
Furthermore, $c_1$ cannot be determined by our dimensional reasoning.
So, we prefer to skip any definite numerical estimate for $\tau$
at this stage; but it is clear that if $c_1$ is not zero, the result
will be extremely different from naive expectations based on Feynman
diagrams in the perturbative theory on flat space.

In conclusion, only the complete non perturbative
quantum theory can fix the lifetime of the graviton.
We notice that a relatively short lifetime could have observable
consequences in the frequency power spectrum of gravitational waves, causing
a degradation of this spectrum towards low frequencies,
a circumstance which could explain why they are so difficult to detect.
In spite of their weakness, gravitational waves could then reveal
quantum gravity effects.

The autor would like to thank D.\ Maison and the staff of MPI for the
kind hospitality in Munich.

\end{document}